\documentclass[a4paper, 12pt]{article}

\usepackage{hyperref}
\usepackage{fancyvrb}
\usepackage{color}
\usepackage{placeins}
\usepackage{framed}
\usepackage[sorting=ydnt,style=authoryear,natbib=true,backend=bibtex]{biblatex}
\bibliography{aging}

\makeatletter
\def\PY@reset{\let\PY@it=\relax \let\PY@bf=\relax%
    \let\PY@ul=\relax \let\PY@tc=\relax%
    \let\PY@bc=\relax \let\PY@ff=\relax}
\def\PY@tok#1{\csname PY@tok@#1\endcsname}
\def\PY@toks#1+{\ifx\relax#1\empty\else%
    \PY@tok{#1}\expandafter\PY@toks\fi}
\def\PY@do#1{\PY@bc{\PY@tc{\PY@ul{%
    \PY@it{\PY@bf{\PY@ff{#1}}}}}}}
\def\PY#1#2{\PY@reset\PY@toks#1+\relax+\PY@do{#2}}

\expandafter\def\csname PY@tok@gd\endcsname{\def\PY@tc##1{\textcolor[rgb]{0.63,0.00,0.00}{##1}}}
\expandafter\def\csname PY@tok@gu\endcsname{\let\PY@bf=\textbf\def\PY@tc##1{\textcolor[rgb]{0.50,0.00,0.50}{##1}}}
\expandafter\def\csname PY@tok@gt\endcsname{\def\PY@tc##1{\textcolor[rgb]{0.00,0.27,0.87}{##1}}}
\expandafter\def\csname PY@tok@gs\endcsname{\let\PY@bf=\textbf}
\expandafter\def\csname PY@tok@gr\endcsname{\def\PY@tc##1{\textcolor[rgb]{1.00,0.00,0.00}{##1}}}
\expandafter\def\csname PY@tok@cm\endcsname{\def\PY@tc##1{\textcolor[rgb]{0.53,0.53,0.53}{##1}}}
\expandafter\def\csname PY@tok@vg\endcsname{\let\PY@bf=\textbf\def\PY@tc##1{\textcolor[rgb]{0.87,0.47,0.00}{##1}}}
\expandafter\def\csname PY@tok@m\endcsname{\let\PY@bf=\textbf\def\PY@tc##1{\textcolor[rgb]{0.40,0.00,0.93}{##1}}}
\expandafter\def\csname PY@tok@mh\endcsname{\let\PY@bf=\textbf\def\PY@tc##1{\textcolor[rgb]{0.00,0.33,0.53}{##1}}}
\expandafter\def\csname PY@tok@cs\endcsname{\let\PY@bf=\textbf\def\PY@tc##1{\textcolor[rgb]{0.80,0.00,0.00}{##1}}}
\expandafter\def\csname PY@tok@ge\endcsname{\let\PY@it=\textit}
\expandafter\def\csname PY@tok@vc\endcsname{\def\PY@tc##1{\textcolor[rgb]{0.20,0.40,0.60}{##1}}}
\expandafter\def\csname PY@tok@il\endcsname{\let\PY@bf=\textbf\def\PY@tc##1{\textcolor[rgb]{0.00,0.00,0.87}{##1}}}
\expandafter\def\csname PY@tok@go\endcsname{\def\PY@tc##1{\textcolor[rgb]{0.53,0.53,0.53}{##1}}}
\expandafter\def\csname PY@tok@cp\endcsname{\def\PY@tc##1{\textcolor[rgb]{0.33,0.47,0.60}{##1}}}
\expandafter\def\csname PY@tok@gi\endcsname{\def\PY@tc##1{\textcolor[rgb]{0.00,0.63,0.00}{##1}}}
\expandafter\def\csname PY@tok@gh\endcsname{\let\PY@bf=\textbf\def\PY@tc##1{\textcolor[rgb]{0.00,0.00,0.50}{##1}}}
\expandafter\def\csname PY@tok@ni\endcsname{\let\PY@bf=\textbf\def\PY@tc##1{\textcolor[rgb]{0.53,0.00,0.00}{##1}}}
\expandafter\def\csname PY@tok@nl\endcsname{\let\PY@bf=\textbf\def\PY@tc##1{\textcolor[rgb]{0.60,0.47,0.00}{##1}}}
\expandafter\def\csname PY@tok@nn\endcsname{\let\PY@bf=\textbf\def\PY@tc##1{\textcolor[rgb]{0.05,0.52,0.71}{##1}}}
\expandafter\def\csname PY@tok@no\endcsname{\let\PY@bf=\textbf\def\PY@tc##1{\textcolor[rgb]{0.00,0.20,0.40}{##1}}}
\expandafter\def\csname PY@tok@na\endcsname{\def\PY@tc##1{\textcolor[rgb]{0.00,0.00,0.80}{##1}}}
\expandafter\def\csname PY@tok@nb\endcsname{\def\PY@tc##1{\textcolor[rgb]{0.00,0.44,0.13}{##1}}}
\expandafter\def\csname PY@tok@nc\endcsname{\let\PY@bf=\textbf\def\PY@tc##1{\textcolor[rgb]{0.73,0.00,0.40}{##1}}}
\expandafter\def\csname PY@tok@nd\endcsname{\let\PY@bf=\textbf\def\PY@tc##1{\textcolor[rgb]{0.33,0.33,0.33}{##1}}}
\expandafter\def\csname PY@tok@ne\endcsname{\let\PY@bf=\textbf\def\PY@tc##1{\textcolor[rgb]{1.00,0.00,0.00}{##1}}}
\expandafter\def\csname PY@tok@nf\endcsname{\let\PY@bf=\textbf\def\PY@tc##1{\textcolor[rgb]{0.00,0.40,0.73}{##1}}}
\expandafter\def\csname PY@tok@si\endcsname{\def\PY@bc##1{\setlength{\fboxsep}{0pt}\colorbox[rgb]{0.93,0.93,0.93}{\strut ##1}}}
\expandafter\def\csname PY@tok@s2\endcsname{\def\PY@bc##1{\setlength{\fboxsep}{0pt}\colorbox[rgb]{1.00,0.94,0.94}{\strut ##1}}}
\expandafter\def\csname PY@tok@vi\endcsname{\def\PY@tc##1{\textcolor[rgb]{0.20,0.20,0.73}{##1}}}
\expandafter\def\csname PY@tok@nt\endcsname{\def\PY@tc##1{\textcolor[rgb]{0.00,0.47,0.00}{##1}}}
\expandafter\def\csname PY@tok@nv\endcsname{\def\PY@tc##1{\textcolor[rgb]{0.60,0.40,0.20}{##1}}}
\expandafter\def\csname PY@tok@s1\endcsname{\def\PY@bc##1{\setlength{\fboxsep}{0pt}\colorbox[rgb]{1.00,0.94,0.94}{\strut ##1}}}
\expandafter\def\csname PY@tok@gp\endcsname{\let\PY@bf=\textbf\def\PY@tc##1{\textcolor[rgb]{0.78,0.36,0.04}{##1}}}
\expandafter\def\csname PY@tok@sh\endcsname{\def\PY@bc##1{\setlength{\fboxsep}{0pt}\colorbox[rgb]{1.00,0.94,0.94}{\strut ##1}}}
\expandafter\def\csname PY@tok@ow\endcsname{\let\PY@bf=\textbf\def\PY@tc##1{\textcolor[rgb]{0.00,0.00,0.00}{##1}}}
\expandafter\def\csname PY@tok@sx\endcsname{\def\PY@tc##1{\textcolor[rgb]{0.87,0.13,0.00}{##1}}\def\PY@bc##1{\setlength{\fboxsep}{0pt}\colorbox[rgb]{1.00,0.94,0.94}{\strut ##1}}}
\expandafter\def\csname PY@tok@bp\endcsname{\def\PY@tc##1{\textcolor[rgb]{0.00,0.44,0.13}{##1}}}
\expandafter\def\csname PY@tok@c1\endcsname{\def\PY@tc##1{\textcolor[rgb]{0.53,0.53,0.53}{##1}}}
\expandafter\def\csname PY@tok@kc\endcsname{\let\PY@bf=\textbf\def\PY@tc##1{\textcolor[rgb]{0.00,0.53,0.00}{##1}}}
\expandafter\def\csname PY@tok@c\endcsname{\def\PY@tc##1{\textcolor[rgb]{0.53,0.53,0.53}{##1}}}
\expandafter\def\csname PY@tok@mf\endcsname{\let\PY@bf=\textbf\def\PY@tc##1{\textcolor[rgb]{0.40,0.00,0.93}{##1}}}
\expandafter\def\csname PY@tok@err\endcsname{\def\PY@tc##1{\textcolor[rgb]{1.00,0.00,0.00}{##1}}\def\PY@bc##1{\setlength{\fboxsep}{0pt}\colorbox[rgb]{1.00,0.67,0.67}{\strut ##1}}}
\expandafter\def\csname PY@tok@kd\endcsname{\let\PY@bf=\textbf\def\PY@tc##1{\textcolor[rgb]{0.00,0.53,0.00}{##1}}}
\expandafter\def\csname PY@tok@ss\endcsname{\def\PY@tc##1{\textcolor[rgb]{0.67,0.40,0.00}{##1}}}
\expandafter\def\csname PY@tok@sr\endcsname{\def\PY@tc##1{\textcolor[rgb]{0.00,0.00,0.00}{##1}}\def\PY@bc##1{\setlength{\fboxsep}{0pt}\colorbox[rgb]{1.00,0.94,1.00}{\strut ##1}}}
\expandafter\def\csname PY@tok@mo\endcsname{\let\PY@bf=\textbf\def\PY@tc##1{\textcolor[rgb]{0.27,0.00,0.93}{##1}}}
\expandafter\def\csname PY@tok@mi\endcsname{\let\PY@bf=\textbf\def\PY@tc##1{\textcolor[rgb]{0.00,0.00,0.87}{##1}}}
\expandafter\def\csname PY@tok@kn\endcsname{\let\PY@bf=\textbf\def\PY@tc##1{\textcolor[rgb]{0.00,0.53,0.00}{##1}}}
\expandafter\def\csname PY@tok@o\endcsname{\def\PY@tc##1{\textcolor[rgb]{0.20,0.20,0.20}{##1}}}
\expandafter\def\csname PY@tok@kr\endcsname{\let\PY@bf=\textbf\def\PY@tc##1{\textcolor[rgb]{0.00,0.53,0.00}{##1}}}
\expandafter\def\csname PY@tok@s\endcsname{\def\PY@bc##1{\setlength{\fboxsep}{0pt}\colorbox[rgb]{1.00,0.94,0.94}{\strut ##1}}}
\expandafter\def\csname PY@tok@kp\endcsname{\let\PY@bf=\textbf\def\PY@tc##1{\textcolor[rgb]{0.00,0.20,0.53}{##1}}}
\expandafter\def\csname PY@tok@w\endcsname{\def\PY@tc##1{\textcolor[rgb]{0.73,0.73,0.73}{##1}}}
\expandafter\def\csname PY@tok@kt\endcsname{\let\PY@bf=\textbf\def\PY@tc##1{\textcolor[rgb]{0.20,0.20,0.60}{##1}}}
\expandafter\def\csname PY@tok@sc\endcsname{\def\PY@tc##1{\textcolor[rgb]{0.00,0.27,0.87}{##1}}}
\expandafter\def\csname PY@tok@sb\endcsname{\def\PY@bc##1{\setlength{\fboxsep}{0pt}\colorbox[rgb]{1.00,0.94,0.94}{\strut ##1}}}
\expandafter\def\csname PY@tok@k\endcsname{\let\PY@bf=\textbf\def\PY@tc##1{\textcolor[rgb]{0.00,0.53,0.00}{##1}}}
\expandafter\def\csname PY@tok@se\endcsname{\let\PY@bf=\textbf\def\PY@tc##1{\textcolor[rgb]{0.40,0.40,0.40}{##1}}\def\PY@bc##1{\setlength{\fboxsep}{0pt}\colorbox[rgb]{1.00,0.94,0.94}{\strut ##1}}}
\expandafter\def\csname PY@tok@sd\endcsname{\def\PY@tc##1{\textcolor[rgb]{0.87,0.27,0.13}{##1}}}

\def\PYZbs{\char`\\}
\def\PYZus{\char`\_}
\def\PYZob{\char`\{}
\def\PYZcb{\char`\}}
\def\PYZca{\char`\^}

\def\PYZlt{\char`\<}
\def\PYZgt{\char`\>}
\def\PYZsh{\char`\#}
\def\PYZpc{\char`\%}
\def\PYZdl{\char`\$}
\def\PYZhy{\char`\-}
\def\PYZsq{\char`\'}
\def\PYZdq{\char`\"}

\makeatother

\usepackage{lineno}
\usepackage{graphicx}

\def\citerr #1 {(\cite {#1})}

\begin{document}

\begin{center}
{\Large \sf 
Aging as a mean to retain an adaptive mutation rate in mutagenesis with asymmetric reproduction}
\end{center}

\abstract{
\bf 
The paper discusses a connection between asymmetric reproduction -- that is reproduction in a parent-child relationship where the parent does not mutate during reproduction --, the fact that all non-viral lifeforms bear genes of their reproduction machinery and how this could relate to evolutionary mechanisms behind aging. In a highly simplified model of the evolution process rules are derived under which aging is an important factor of the adaption in the evolution process and what groups of life-forms necessarily have to age and where exceptions from that rule are possible.}

‪


\newpage

\section*{\sf Variables \& Constants}
\begin {tabular} [t] {cl}
${\cal I}_{i,t}=[e,m]$   & Individual with id $i$ and generation number $t$ \\
$e$			& Individual's guess of the target value \\
$m$		        & Individual's mutation rate \\
${\cal F}_i $		& Fitness of an individual \\
${\cal F}_{c,i} $	& Fitness including childhood impairment \\
${\cal F}_{a,i} $	& Fitness including childhood impairment and aging \\
${\cal I}_{*,i,t+1}$ & Individual after mutating from generation $t$, but before selection \\
$[{\cal I}]_t$     & Total population of one generation \\
$ \tau$            & Target value of evolutionary optimization\\
$\nu_{0,\sigma}$        & Normal distributed zero mean random number with $\sigma$ standard variance \\
$\cal M$	& Mutation operator \\
$\cal S_\tau $  & Selection operator
\end{tabular}

\newpage

\section{Introduction}

Most forms of life base on an evolutionary process that not only covers information about reproduction, but also about the reproduction machinery itself. Due to the nature of biological encoding thus the reproduction machinery is subject to some kind of variability which also effects the accuracy and mutation rate of this machinery. In other words the genome includes an encoding of the mutation rate, the mutation rate is subject to mutation itself. Neither does a perfectly accurate reproduction nor an extremely error prone reproduction machinery optimally serve the purpose of survival of the species. Rather  it seems that the mutation rate is itself result of an optimization process that relates to the environment or state of the host. 

Life forms that do not include information about their reproduction machinery, thus rely on their host, are viruses. However, usually obstinate viruses bring some gene with them  that is capable to influence their mutation rate. So, this feature covers a crucial advantage which appears to be very important to maintain for both primitive and all higher forms of life. 

Although information about the reproduction machine is readily included into higher life forms it is not sure that mechanism still works under all circumstances. The reason is that parents and offspring coexist after reproduction and the parents remain not-mutated during the reproduction. 
Usually the offspring is not competitive during childhood and thus has a disadvantage against
their parents, which on the long run limits the adaptiveness of mutation rates.
In the following we argue that a preprogrammed limited lifespan -- aging -- is exactly a mean a to ensure that advantages of an adaptive mutation rate. Thus, the reason for human aging is to keep the mutation rate adaptive.

A simple numerical model that is outlined below illustrates this advantage. It can be shown that in a some kind of static environment the mutual information between the target environment and the adaptive individual life form is continuously increased during the evolutionary process. 

We try to model evolution as an adaptive optimization process, reinforcement learning and supervised learning, where the feet-back of the fitness function is blurred by a stochastic component. One aim of that paper is to show that in the case of a static landscape of the cost function the mutual information of the optimal value and the population is continuously increasing if the mutation rate can be controlled by the genome. However, the process is limited if the mutation rate is not part of the process. The underlying mechanism is a common technique in evolutionary algorithms (see for example \citerr {banzhaf2010} ). The impact on biological evolution has been discussed and seems to be somewhat generally accepted in the community 
\citerr {2008pigliucci} . 

In a second stage the paper proposes a connection to another phenomenon of higher life: that is aging. Here it is aimed to indicate that aging and predetermined life span is a necessary requirement to guarantee an continuous improvement and refinement of the optimization process.

\subsection{Aging}

Various explanations for aging in higher forms of life compete \citerr {whydoweage2000, evolutionofaging2013, systemsbiology2011} . The probably oldest explanation was by Weismann  in the nineteenth century who proposed that aging exists to ‘make room for the young’ (The author found the quote in \citerr {evolutionofaging1977} .).

Next to the suggestion that aging could be the result of some kind of wearing out of organs, cells, muscles, bones, tendons etc. (see for example \citerr {phenptosis2011} ), is one dominant idea with regard to aging. Closely related was also the idea that bio-mechanics at cell level necessarily lead to vanishing fertility at higher ages which makes older individuals useless for the gene pool.

In addition, other ideas exist such as to assume social functions in aging \citerr {chaotic2006,2014mittendorf} . Here one may object that aging is part of life various types of life including animals with very different social behaviors and societies.

An open question is why some species apparently do not age \citerr {treeoflife2012} .  However, this argument has been refuted by the example of {\em Gopherus agassizii}, a desert turtle, that shows a reduced mortality a higher ages and also no decrease of fertility then, i.e. they are not aging \citerr {rootofaging2012} . 

Technically wearing out of course is an intrinsic part of aging. However, nature may come up with solutions, i.e. the replacement of worn out parts as it is common practice for many types of cells in the human body. There is no reason to assume that a principle difference between the metabolisms of human and these desert turtles necessarily requires aging in humans while it is not necessary for these animals. One important observation is that while early aging can be a result of genetic disease (i.e. progeria), no cases of humans are known that due to a genetic "defect" or other reasons do not age -- while longevity in fact is genetically determined. Finally it is to note that aging is the most important life limiting factor in humans. Extrapolating annual death rates at the age of 30--35 would lead to a life expectancy higher than 250 years\footnote{Assuming an annual mortality smaller than 0.4\%}.

The basic idea is here to proof that aging, i.e. the systematic eradication of a generation after a certain period of time is the result of a process that tries to control the mutation rate and thus in a way the trade off between the information transfer about the environment between two generations and the potential improvement that is done by the mutation rate.

\begin{figure}[t]
\begin{center}
\centerline{\includegraphics[width=14.5cm]{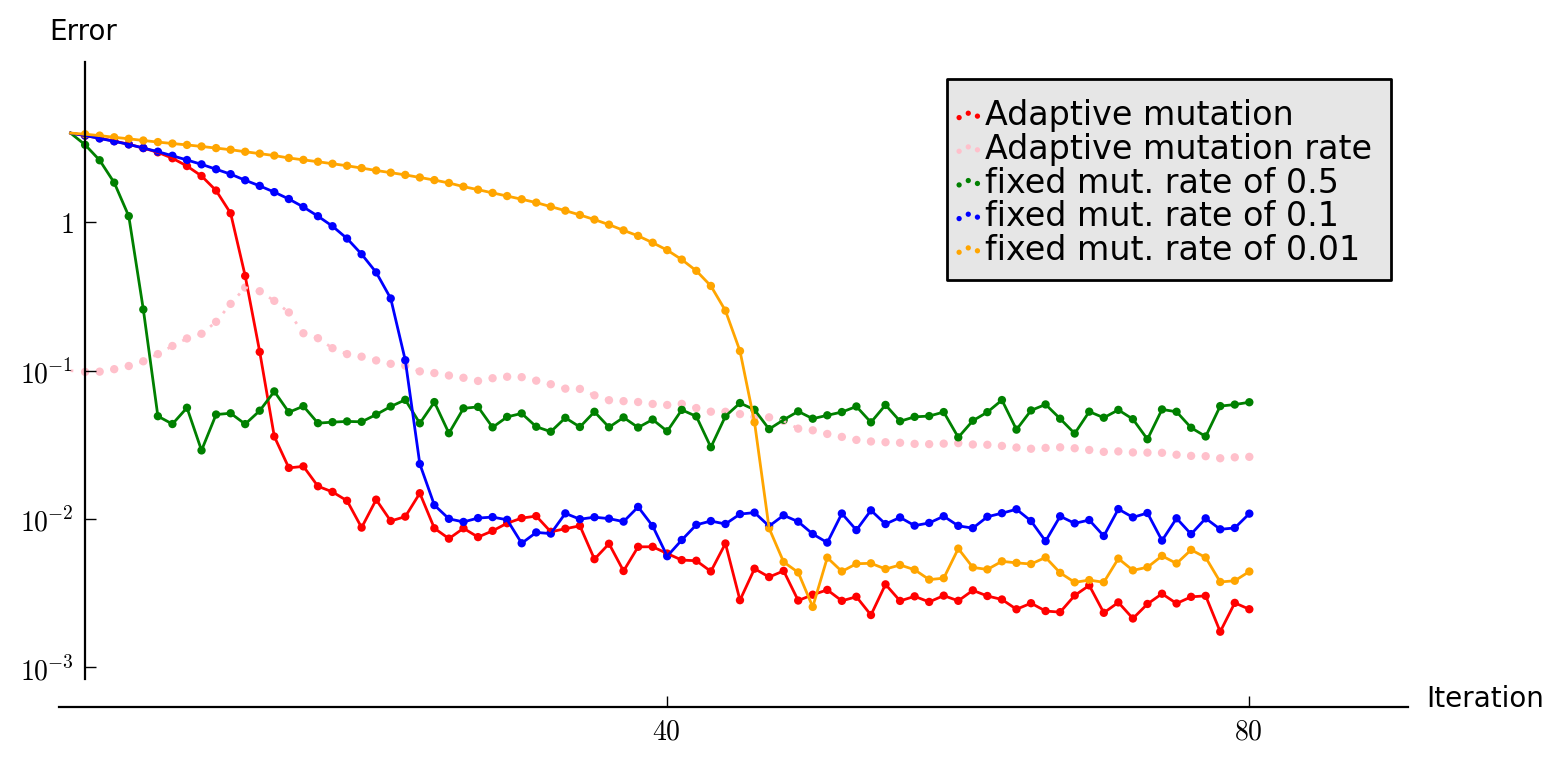}}

\caption{Development of individuals according to the model of sect. \ref{simplest_model} in comparison to an evolution process (red, mutation rate: pink) with evolutions using fixed mutation rates.
Simulation details: In each case 60 individuals, 80 generations, all evolutions start at around 1.0, fitness is potential with a single minimal point at $\tau=5.0$, noise level of the observation
($\sigma=0.0$). 
\label{develop00}
}

\end{center}
\end{figure}

\section{Simplest evolutionary model}

\subsection{Simplest individuals that include information about reproduction error\label{simplest_model}}

We see evolution as an optimization process on a fitness function, which in the simplest case only depends on one scalar parameter $e$ that is unknown to the individual and is guessed during the process. 
In addition, we have a second parameter  $m$ which is a measure of the mutation rate. We suggest the following sequence of processes during one generation:

\begin{itemize}
\item The tupel
\begin{equation}
{\cal I}_{i,t} = [e,m]_{i,t} \in R^{2},
\end{equation} 
shall represent an individual, the vector $[{\cal I}]_t$ the whole population of $N$ surviving individuals of one generation. 

\item Inheritance and Mutation from one generation to the next can be expressed by 
\begin{equation}
{\cal I}_{*,i,t+1} = {\cal M} \, {\cal I}_i = [e_i + m_i \, \nu_{0,1} \, , \, m_i \, (1+m_i \nu_{0,1})],
\end{equation} 
where $\nu_{0,1}$ is a zero mean normal random value of a distribution with variance 1.
\item Selection process (in the following the symbolic operator ${\cal S}_{\bf t}$) can be accessed with regard to a target value $\tau$ that has to be estimated by each individual:
\begin{equation}
[{\cal I}]_{t+1} = {\cal S}_{\tau} {\cal I}_{*,i, t+1} = [ {\cal I}_{*,i^*, t+1}], \; \; {\rm where} \; \; i^* \in {\arg \max}_{i} {\cal F}_i,
\end{equation}
where $\arg \max$ produces a list of the $k$ items with the largest fitness (i.e. smallest error) values. 
\end{itemize}
In addition, it is assumed that during selection the process ${\cal S}_{T, \sigma}$ the target value can only be observed with an unbiased normal error $\tau=T+\nu_{0,\sigma}$, where if not mentioned otherwise $\sigma$ is set to $0$.
For sake of simplicity we assume:
\begin{equation}
{\cal F}_i = - |e_i - \tau|.
\label{error_func}
\end{equation}

\subsection{Results of the initial simplest model\label{simplest_model_results}}

The model easily reveals that individuals that include information about their gene pool show a better approximation of the target value than individuals with fixed mutation rates. 
Figure \ref{develop00} gives an impression of such numerical experiments. One can see that the evolution in individuals that bear information about their mutation rate leads to a process where roughly 2 stages can be distinguished:
\begin{itemize}
\item In a first phase the average mutation rate is increasing, the mean of the estimates of the target value is rapidly shifting towards the target value. The improvement of the error is comparable 
to an evolution with a fixed high mutation rate.
\item The second phase the target value is roughly reached further improvements are achieved by the continuous reduction of the mutation rate. Finally, better approximations are reached than for any evolution process using a fixed mutation rate.   
\end{itemize}

\subsection{Information theoretic view on the evolutionary process with adaptive mutation rate}

\begin{figure}[t]
\begin{center}
\centerline{\includegraphics[width=14.5cm]{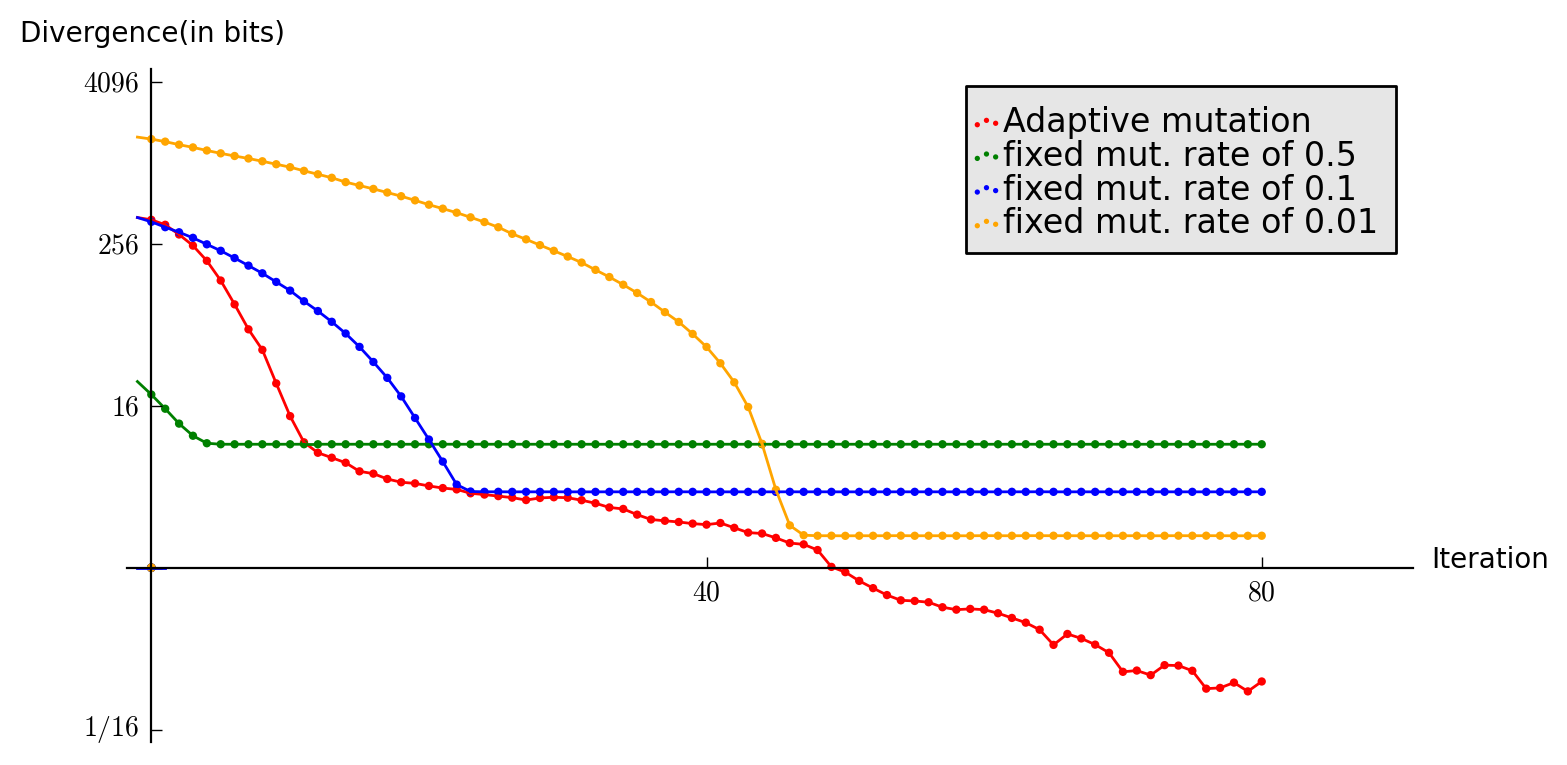}}

\caption{Simulation under the same condition as for fig. \ref{develop00}. Instead of the 
means square error the Kullback-Leibler divergence to the target value with a target error of 
0.02 is calculated.
\label{info}
}

\end{center}
\end{figure}

So far the performance of the individuals with regard to the target have only be discussed in terms of a mean square error. Equivalently, one can use the terminology of information theory and measure the information content of each individual or the complete population with regard to the environment (here the target value). One way to do that is the self-information 
\begin{equation}
I(T) = D_{\mathrm{KL}}(\delta_{T,x} \| \{ p(x) \}), \label{selfinfo1}
\end{equation}
where $\{ p(x) \}$ is a distribution that can be derived from the individuals of the species and 
$D_{\mathrm{KL}}$ is the Kullback-Leibler divergence and $\delta_{T,x}$ is the Dirac function.
Eq. \ref{selfinfo1} measures how much information is missing if one expresses $T$ by means of 
of distribution $p(x)$. However,
the result of eq. \ref{selfinfo1} and say a Gaussian distribution with non-zero variance is always infinity, because an infinitely long adaptation process is necessary until the real valued variable $T$ is determined with infinite accuracy. Thus, it makes sense to settle when $T$ has been found with finite accuracy which can be done by defining
\begin{equation}
I_\sigma(T) = D_{\mathrm{KL}}(G_{T,\sigma,x} \| \{ p(x) \}), \label{selfinfo2}
\end{equation}
where $G(T,\sigma,x)$ is a Gaussian distribution with a mean at $T$ and $\sigma$ standard deviation.
The standard deviation indicates an accuracy that can be considered sufficient and should be reasonable small. For $p(x)$ it makes sense also to assume a Gaussian,
\begin{equation}
p(x) = G_{e,m,x},
\end{equation} 
where $e$ and $m$ are estimate and mutation rate of an individual of the population. Thus, the equation
\begin{equation}
I_\sigma(T) = D_{\mathrm{KL}}(G_{T,\sigma,x} \| G_{e,m,x} ), \label{selfinfo3}
\end{equation}
where $e$ is the estimate and $m$ is the mutation rate of the individual.
Thus, in the outlined sense eq. \ref{selfinfo3} can give a measure of the 
on how accurately the individual is adapted to the environment. 
Fig. \ref{info} depicts results for a simulation that is analogue to the simulation depicted in fig.
\ref{develop00}. 

One can see that a species with a adaptive mutagenesis outpaces the performances of all species with fixed mutation rates. The evolutionary process is scaling automatically scaling up the performance and thus aquires better and better knowledge of the environment, whereas the performance of the species with the fixed mutation rates go into saturation after a while.

The conclusion in this work is that
strategies to guarantee a control of the mutation are an important feature to improve performance 
in evolution.
They may vary for different life forms, for example:
Retro-viruses have a copying mechanism from RNA to DNA that precesses the actual 
reproduction. The copying is done by the Reverse Transcriptase that comes with the genome of the 
virus. Thus, retro-viruses not only have a high but also at least potentially an adaptive 
mutation rate.

\begin{figure}[t]
\begin{center}
\centerline{\includegraphics[width=14.5cm]{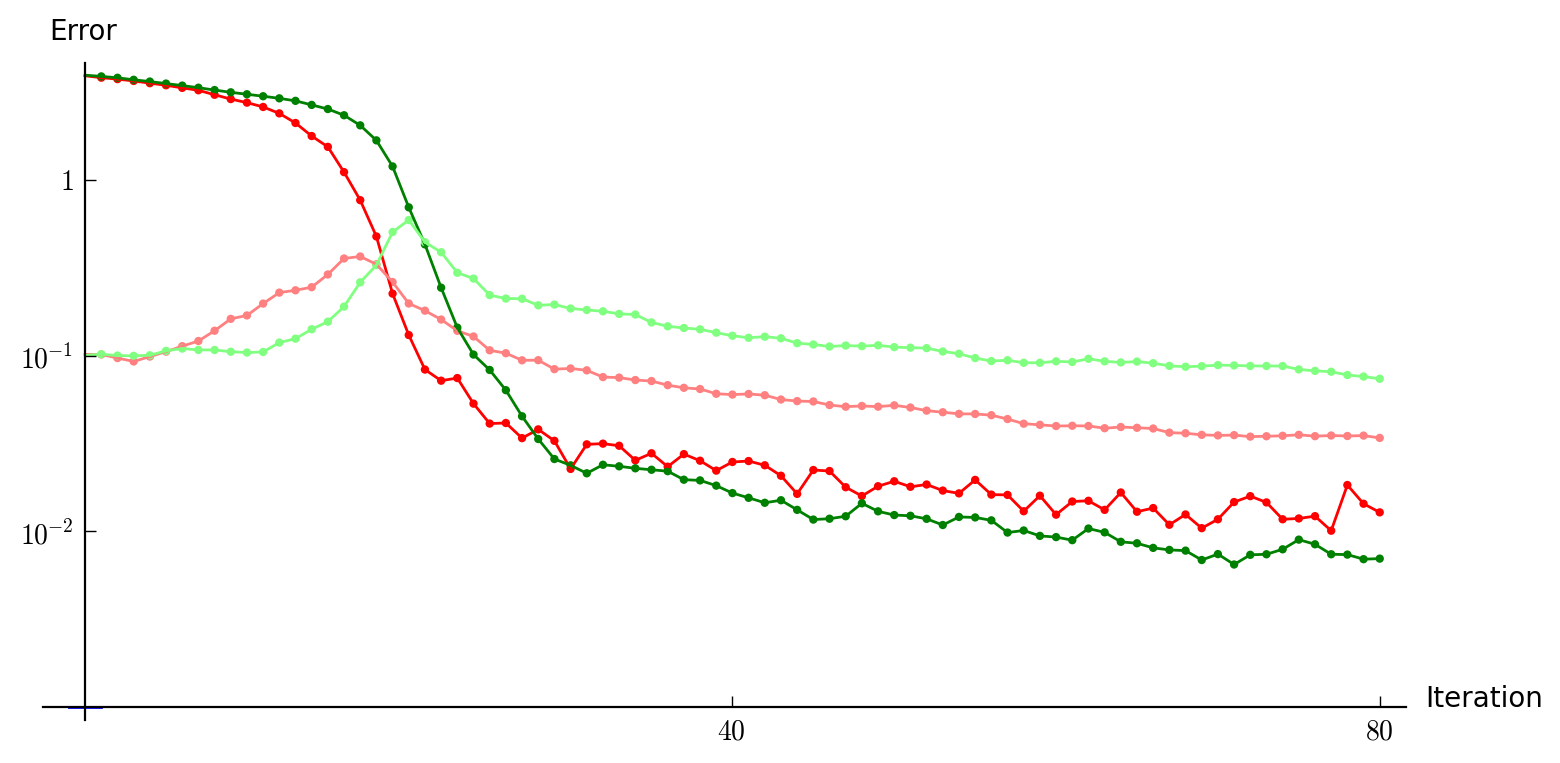}}

\caption{A comparison between the development process between the symmetric (average error from target value in dark red) and asymmetric (average error in dark green) reproduction scenario 
(see sect. \ref{assym_model}).  Also plotted are the average mutation rates (in light red and light green respectively).
The initial asymmetric model does not have an overall significant worse performance than the symmetric model, although for an identical initial mutation rate the convergence to the target value is a bit slower in the asymmetric case in comparison to the symmetric case. 
\label{develop01}
}

\end{center}
\end{figure}

\subsection{Asymmetric reproduction scenario\label{assym_model}}

In the following we would like to discuss effects of reproduction if one site of the duplication process mutates and the other side does not mutate.
From evolutionary point of view that circumstance separates parents from offspring.
This is true for all kind of parent -- offspring relationships, that includes of course higher animals and humans\footnote{It is not related in any way to sexual reproduction, neither is sexual nor asexual reproduction required nor is it excluded.} 
Fig. \ref{develop01} shows development under such an asymmetric reproduction scheme in comparison to the initial scenario (in Fig. \ref{develop00}). One can see that although the initial adaptation of the asymmetric reproduction scenario is a bit slower in the adaptation process, over a longer adaptation time, the asymmetric reproduction reaches again the performance of the symmetric reproduction scenario, even outruns it over a longer reproduction time. One can also see that the formal adaptation rate of the asymmetric scenario stays higher but on the long run that has no impairing effect on the performance.

\begin{figure}[t]
\begin{center}
\centerline{\includegraphics[width=14.5cm]{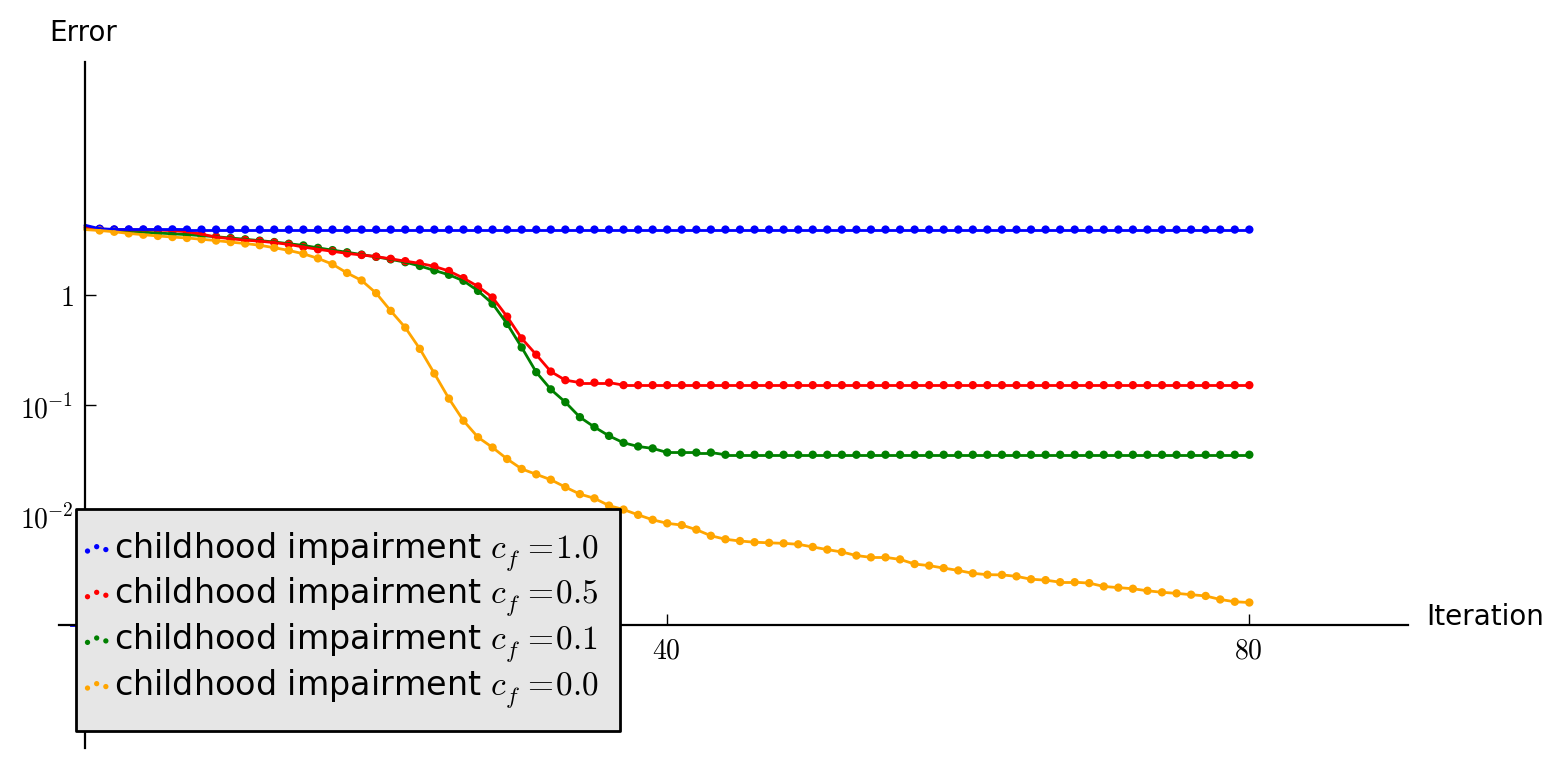}}

\caption{A plot that depicts the impact of different levels of the childhood impairment factor on the development of a species. For our model a level of 1.0 completely prevents the evolutionary process if the process starts from a value around $1.0$ and the target value is $5.0$. 
Lower levels result in an stop of the evolutionary process at a certain value of approximation. 
\label{fig_reduced_fitness}
}

\end{center}
\end{figure}

\subsection{Reduced fitness during early development \label{child_imp}}

It is fair to assume that offspring undergoes some kind of early development. During that time the fitness of the offspring is lacking behind the fitness of the parent. For sake of simplicity it is assumed that the reduced fitness is a constant value which is subtracted from the fitness of adulthood. The value is chosen in a certain range such that reproduction is not prevented completely but still sufficiently strong to see a significant higher death rate than during adulthood (age $>= 1$). Model is thus extended by adding a childhood term to the fitness function:
\begin{equation}
{\cal F}_{c,i} = {\cal F}_i - c_f \times (age_i==0),
\end{equation}
where the fitness ${\cal F}_{c,i}$ replaces the error value from eqn. \ref{error_func}, a lower fitness value results in higher fitness here. $==$ represents the logical operator resulting in $1$ if true and $0$ elsewise. 

Figure \ref{fig_reduced_fitness} shows how different levels of child impairment affect the evolutionary process. The higher the impairment the earlier the development process is frozen in a similar way as we have seen that for fixed mutation rates in Figure \ref{develop00}. After the limit is reached the age of the population increases continuously.

\begin{figure}[t]
\begin{center}
\centerline{\includegraphics[width=14.5cm]{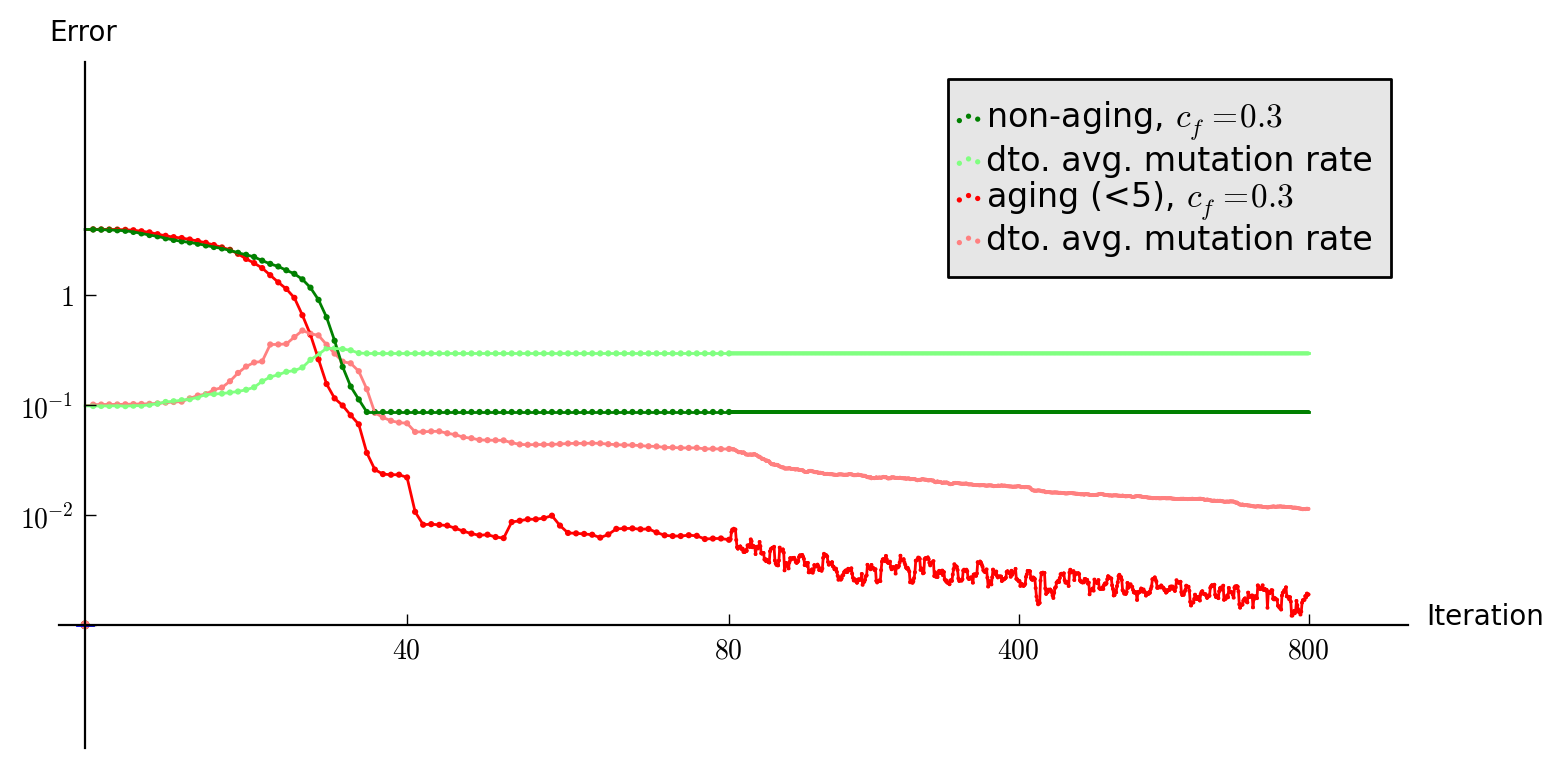}}
\caption{ 
Evolution process of 2 species where one is aging and the other one is non-aging, in both cases a childhood impairment is assumed. The graph shows in detail the first 80 iterations and is then extended to the 800th iteration, where the range from the 80th to the 800th iteration is plotted in the same horizontal range at which the first 80 iterations are depicted. In the case of the aging species extension shows a slow but exponential convergence towards the target value.
\label{aging1_fig}
}
\end{center}
\end{figure}

\subsection{Aging}
It is worth while testing if aging can overcome the limit that origins from the childhood impairment. Thus simulations compare an aging species with a non-aging species where both suffer from the childhood impairment of sect. \ref{child_imp}. For sake of simplicity the aging is introduced in that way that all individuals die at latest before the 6th iteration of life, where the fitness value is increased by a large aging factor $a_f$ that makes further competition impossible. 

\begin{equation}
{\cal F}_{a,i} = {\cal F}_i - c_f \times (age_i==0) - a_f \times (age_i>5),
\end{equation} 
where $age_i<5$ results again in a boolean expression and is $1$ if true.

See Figure \ref{aging1_fig} that depict the results of this simulation. 
Initially (for the first 30 or so iterations) both species show a very similar performance.
However, one can see that the aging species overcomes the limit that was induced by the childhood impairment and continues to approximate exponentially the target value.

\begin{figure}[t]
\begin{center}
\centerline{a. \includegraphics[height=3.8cm]{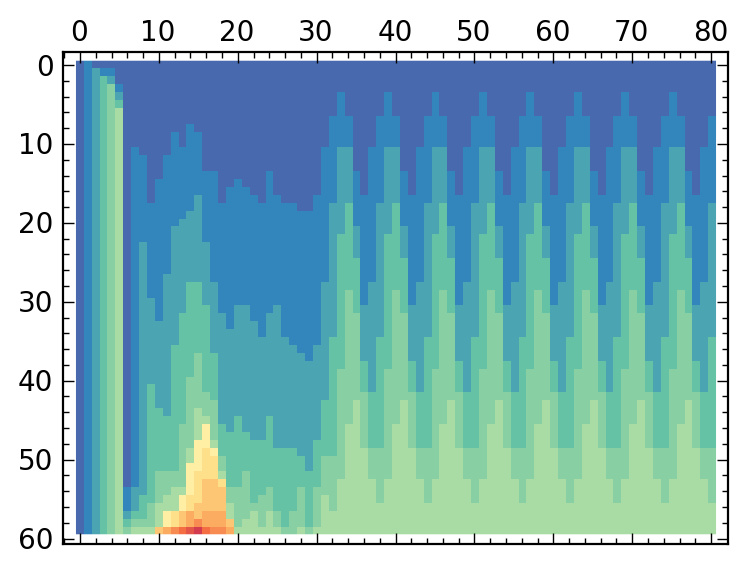}
b. \includegraphics[height=3.8cm]{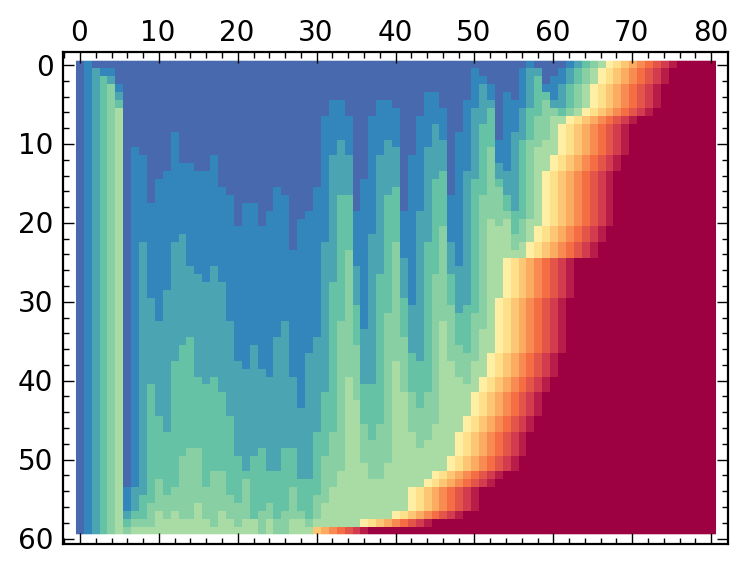}
c. \includegraphics[height=3.8cm]{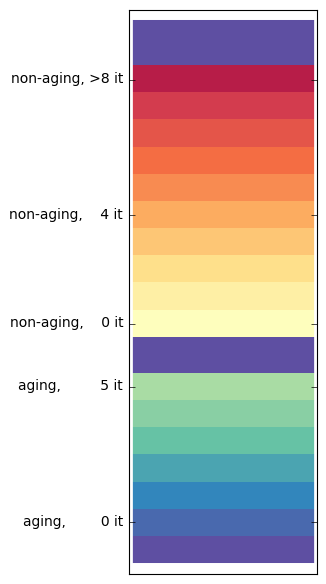}}

\caption{Depicted are cases when one individual within a species that ages is allowed for eternal youth. 
a. Shows an exemplary population dynamics when one individual mutates to eternal youth at the 10th iteration; in b. such a mutations happens in the 30th iteration. Each graph shows the ages of all individuals in each generation and if they belong to the aging or the eternal youth type of the population. X-axis shows the generations and along the Y-axis the individuals are depicted in the manner that they are sorted according to their age. c. shows a scale that decodes the color, according to age and population type. 
Simulation experiments show those genetic mutation tend to die out if the mutation happens during the drift phase of the evolution process, whereas in almost all cases they prevail and finally dominate the population during the convergence phase.  
}
\label{population_switch}
\end{center}
\end{figure}

\begin{figure}[t]
\begin{center}
\centerline{\includegraphics[width=14.5cm]{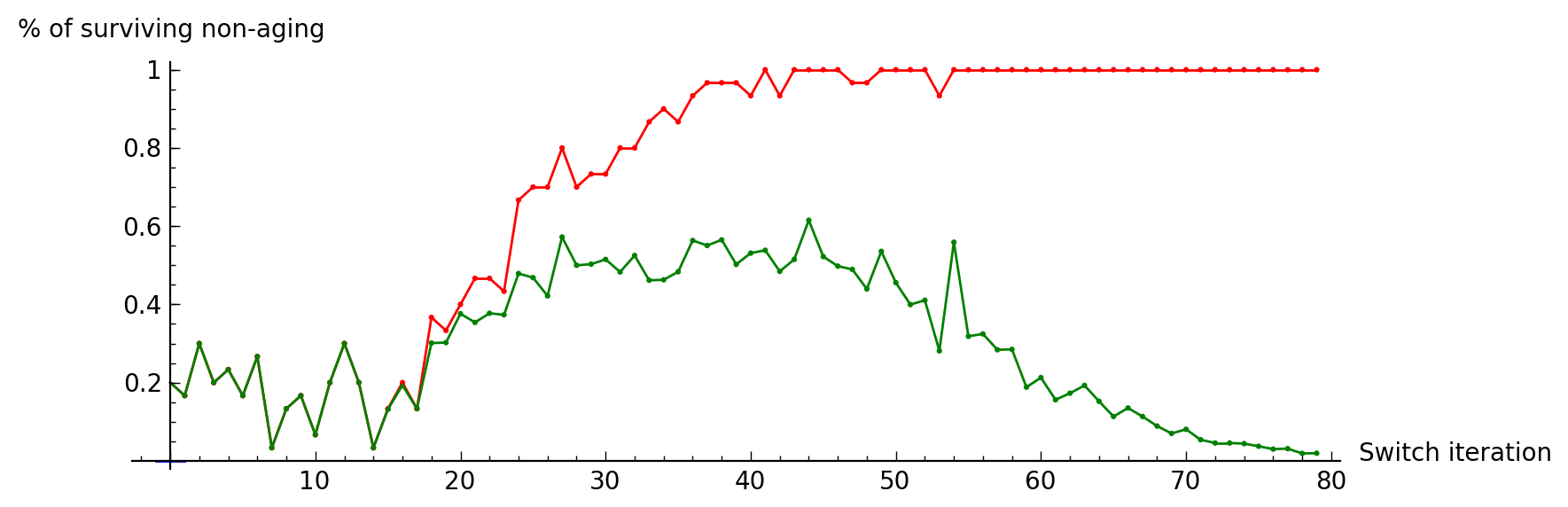}}
\caption{ Statistic of which ratio of mutations from aging to non-aging survive until the end of the simulation at iteration 81. 
The x-axis depicts the iteration at which exactly one individual of the population switches from aging to non-aging. At the end of the simulation usually all individuals are either of the aging or the non-aging type, but also mixed populations of both types may occur. 
The red curve shows what ratio of the populations contain at least one individual of the non-aging type, the green line shows what ratio of individuals is non-aging on average at the end of the simulation. The final decay of the green curve is caused by the fact that the non-aging part for the population was still proliferating when the simulation stopped.  The graph was derived as an average of 30 simulations for each point of the graph. See 
\label{aging6_fig}
}
\end{center}
\end{figure}

\subsection{Evolution with an aging gene that can be turned on and off}
In the following it is assumed that aging is an acquired feature of the evolutionary process. 
In the framework of the model one can test how mutations from aging to eternal youth and from eternal youth to aging are related to
\begin{itemize}
\item the relative fitness of individuals with and without aging within the population and 
\item the absolute average fitness of the whole population during the evolutionary adaptation process.
\end{itemize}
Thus, if can be tested if an non-aging gene set successfully can compete and survive in an aging population, and if so how does this affect the evolutionary process of the whole population.  
Figure \ref{population_switch} shows examples for cases when at certain iterations one of the individuals is switched from aging to eternal youth at one time of the development. These 2 examples are typical for the development. While a switch from aging to non-aging during the early phase of the evolutionary process, that is when the average of the estimate is still drifting and the mutation rate is increasing, the non-aging individual strain usually dies out, with a few exceptions. At later stages of the development the picture is completely different: Here virtually in all cases a mutation from non-aging to aging survives (see also Figure \ref{aging6_fig}), proliferates and finally gets over the complete population. The reason for this is that the number surviving children in a population with a constant number of individuals is basically determined by the number of individuals whose life is completed due to the aging process. Thus, children are superseded more and more by non-aging individuals. For the overall population the result is that the evolution is frozen at the given state of the development. If the aging population persists one can see a regular pattern emerge which is caused by the fact that in every generation only those individuals are replaced by offspring who have reached the limit of their lifespan. Thus, the pattern in this case has exactly the same periodicity as as the limit of life (in this case 5 iterations). 

Figure \ref{aging6_fig} gives an overview of the statistics of how strong is the relative fitness of the non-aging gene in a species that initially does age. While non-aging genotypes in most cases die out during an early stage of the evolutionary process, it is --due to the effects described above-- a very fit mutation at the later stages of the development.

\section{Conclusions and predictions drawn from the model}

Overall we see the following impact of aging on the evolutionary process:
Aging in the model helps to overcome the childhood impairment and thus --to put it in the words of Weismann -- that the old have to die in order to make space for the younger. 
In terms of evolution in our simplistic model: One can distinguish the early stage of the development where the individuals still show an increasing mutation rate. During this phase non-aging genes die out within the species, the aging gene is more fit than the non-aging gene. 
At later stages of the development the simulation shows a continuous reduction of the mutation rate. If during that phase individuals appear in one species that do not age anymore the adaptation process of the whole species is interrupted. In the following the species remains on the same level of evolution. Thus, in the competition of species the species that allow non-aging individuals have to die out, if the evolutionary process has not reached a final static state. Thus, species would tend to develop a hard barrier against the appearance of non-aging genes in order preserve the ongoing process of evolution. Finally in cases of a species that has survived over a long time in one niche without larger changes in their environment, non-aging may be acceptable because evolutionary adaptation has ended almost and the pressure from other species is negligible. Thus one may derive the new rules to predict if a species is aging or non-aging.
Thus, a species using asymmetric reproduction would age, except
it has survived in a highly specialized niche for much time in evolutionary history and is not under evolutionary pressure.

 \printbibliography 
\end{document}